\documentclass{aa}
\usepackage{graphicx}
\newcommand{\ltsima} {$\; \buildrel < \over \sim \;$}
\newcommand{\gtsima} {$\; \buildrel > \over \sim \;$}
\newcommand{\lta} {\lower.5ex\hbox{\ltsima}}
\newcommand{\gta} {\lower.5ex\hbox{\gtsima}}
\begin{document}
%\thesaurus{03}
\title{Are radio galaxies and quiescent  galaxies different? Results from the analysis of HST brightness profiles.
\thanks{Based on observations with the NASA/ESA Hubble Space Telescope obtained at the
Space Telescope Science Institute, which is operated by AURA, Inc.,
under NASA contract NAS 5-26555 and by STScI grant GO-3594.01-91A}}
\author{H.R. de Ruiter\inst{1,2} \and P. Parma\inst{2} \and A. Capetti\inst{3}
\and R. Fanti\inst{4,2} \and R. Morganti\inst{5} \and L. Santantonio\inst{6}
}
\institute{INAF - Osservatorio Astronomico di Bologna, Via Ranzani, 1, I-40127 Italy
\and INAF - Istituto di Radioastronomia, Via Gobetti 101, I-40129, Bologna, Italy
\and INAF - Osservatorio Astronomico di Torino, Strada Osservatorio 25,  I-10025 Pino Torinese, Italy
\and Istituto di Fisica, Universit{\`a} degli Studi di Bologna, Via Irnerio 46, I-40126 Bologna, Italy
\and Netherlands Foundation for Research in Astronomy, Postbus 2, NL-7990 AA, Dwingeloo, The Netherlands
\and Universit{\`a} degli Studi di Torino, Via Giuria 1, I-10125 Torino, Italy
}
\offprints{H.R.\, de Ruiter}
\date{Received; accepted}
\titlerunning{Radial Brightness Profiles of B2 Radio Galaxies}
\authorrunning{de Ruiter et al.}
\maketitle

\begin{abstract}
\label{abstract}

We present a study of the optical brightness profiles of early type galaxies, using a number of samples of radio galaxies
and optically selected elliptical galaxies. For the radio galaxy samples--B2 of Fanaroff-Riley type I and 3C of Fanaroff-Riley type II-- we determined
a number of parameters that describe a "Nuker-law" profile, which were compared with those already known for the optically selected objects.

We find that radio active galaxies are always of the "core" type (i.e. an inner Nuker law slope $\gamma < 0.3$). However, there are core-type galaxies which harbor no significant radio source and which are indistinguishable from the radio active galaxies. We do not find any
radio detected galaxy with a power law profile ($\gamma > 0.5$). This difference is not due to any effect with absolute magnitude, since 
in a region of overlap in magnitude the dichotomy between radio active and radio quiescent galaxies remains. We speculate that core-type
objects represent the galaxies that have been, are, or may become, radio active at some stage in their lives; active and non-active core-type 
galaxies are therefore identical in all respects except their eventual radio-activity: on HST scales we do not find any relationship between
boxiness and radio-activity.

There is a fundamental plane, defined by the parameters of the core (break radius $r_b$ and break brightness $\mu_b$), which
is seen in the strong correlation between $r_b$ and $\mu_b$. The break radius is also linearly proportional to the optical Luminosity
in the $I$ band. Moreover, for the few galaxies with an independently measured black hole mass, the break radius turns out to be
tightly correlated with $M_{BH}$. The black hole mass correlates even better with the combination of fundamental plane 
parameters $r_b$ and $\mu_b$, which represents the central velocity dispersion.

\end{abstract}
\section{Introduction}
\label{sec:intro}

One of the unanswered problems in astronomy is why some (early-type) galaxies produce powerful radio sources and
others do not. Attempts to find optical signatures of the process of radio source formation have usually failed, although some
hints have emerged that the general activity of galactic nuclei is linked to the merging of galaxies and the consequent re-organization
of the nuclear region (Heckman et al. \cite{Hec86}; Wilson \& Colbert \cite{Wil95}).

The Hubble Space Telescope (HST) has opened up the possibility of studying the nuclear regions of galaxies in great detail, which resulted in a
number of studies concerning the nuclear properties of a variety of objects. From the by now vast literature we give just a few examples: nearby elliptical galaxies (Crane et al. \cite{Cra93}; Jaffe et al. \cite{Jaf94}; Faber et al. \cite{Fab97}; Verdoes Kleijn et al. \cite{Ver99}), 
brightest cluster members (Laine et al. \cite{Lai03}), and objects at higher redshifts (McLure et al. \cite{Mcl04}).

It was thus found that the inner profile may contain a central cusp at small radii, in contrast to the previously accepted idea of a constant 
density core.  The profiles can be roughly divided into two types: ones with a "break" radius at which the slope of the profile
changes and becomes shallower inwards, and those with at most a small change in slope. The latter come under the name
of power-law types and are common among the weaker galaxies with $M_v > -20.5$ ($H_0 = 80$\ km\ s$^{-1}$Mpc$^{-1}$); the former, called core-type galaxies (see Lauer et al. \cite{Lau95}; Jaffe et al. \cite{Jaf94}), are more common among the brighter galaxies ($M_v < -22$, Faber et al. \cite{Fab97}).

Several observables are known to be correlated, notably the central velocity dispersion, the effective radius, and its corresponding surface
brightness (Djorgovsky \& Davis \cite{Djo87}; Dressler et al. \cite{Dre87}). Together they describe a linear plane, known as the fundamental
plane of elliptical galaxies (see e.g. Bettoni et al. \cite{Bet01}; Woo et al. \cite{Woo04}). In turn these parameters, or a 
combination of them, correlate with the mass of the central black hole. The correlation between central velocity dispersion and black hole mass was first found by Ferrarese \& Merritt (\cite{Fer00}) and Gebhardt et al. (\cite{Geb00}); further discussion can be found in, e.g., Graham et al. (\cite{Gra01}); Erwin et al. (\cite{Erw04}).  With the
help of these relations black hole masses have been determined for several samples of radio galaxies (Bettoni et al. \cite{Bet03}; McLure et al. \cite{Mcl04}).
The parameters making up the fundamental plane are, in a sense, global (at least the effective radius and surface brightness). However, 
with the parameters that are more related to the conditions in the nucleus, like break radius $r_b$ and break surface brightness $\mu_b$, 
we can form a similar "core" fundamental plane (Faber et al. \cite{Fab97}). Since there are some plausible theoretical reasons why the inner profiles of the galaxies are influenced by the presence of a central black hole, (e.g. van der Marel \cite{Van99}; Ravindranath et al. \cite{Rav02}; Milosavljevi{\'c} et al. \cite{Mil02}), a careful study of the inner profiles may reveal important 
information on the physical conditions in the centers of galaxies. 

We will discuss this point in the light of a detailed analysis of existing HST data listed in the next sections. In particular we study the possible differences between galaxies with an AGN (in our case a radio emitting nucleus) and non-active ellipticals, using
i) low luminosity (FR type I) radio sources, selected from the B2 sample (Fanti et al. \cite{Fan87}), ii) powerful nearby 3C radio galaxies of type FR-II, iii) nearby "normal" ellipticals, which are not necessarily radio emitters,
taken from the sample described by Faber et al. (\cite{Fab97}), and finally, iv) galaxies that are the brightest cluster members (the sample discussed by Laine et al. \cite{Lai03}).
HST's high resolution images in the $F555W$ (similar to $V$) and $F814W$ (similar to $I$) filters, taken with exposure times of 300 seconds, exist for a majority (57/100)  of B2 radio galaxies (Capetti et al. \cite{Cap00}; de Ruiter et al. \cite{Der02}). These provide an excellent starting 
point for studying possible relations between radio and optical properties, so we discuss these data in some detail. 
Images of Fanaroff-Riley type II (3C) radio galaxies were collected from the HST archive. For comparison purposes we
used data from a study of brightest cluster galaxies (Laine et al. \cite{Lai03}) and from the analysis of
elliptical galaxies done by Faber et al. (\cite{Fab97}).

In Sect. \ref{sec:thesample} we briefly describe the HST observations of B2 radio galaxies, the selection criteria of the
3C FR-II radio galaxies, and the method of
fitting brightness profiles. In Sect. \ref{sec:discussion} we first discuss the properties of the brightness profile
fits, followed by an analysis of the possible correlations of radio and optical properties. Finally in Sect. \ref{sec:conclusions} we give a
summary of our conclusions. In order to facilitate comparison with other articles on the subject,
all intrinsic parameters (radio power, absolute magnitudes, sizes) were calculated with
$H_0 = 80$ km s$^{-1}$Mpc$^{-1}$.

\section{Selection of the data and profile fitting}
\label{sec:thesample}

\subsection{The data}
\label{subs:data}
Only part of the B2 sample of radio galaxies has been observed with the HST (see previous section). The 57 objects with HST data are a random selection of the total B2 sample and should therefore constitute an unbiased subsample.
An extensive description of the HST observations and of the data reduction steps was
given in Capetti et al. (\cite{Cap00}), so we refer to this article for further details on the HST imaging of B2 galaxies. The properties of circum-nuclear dust were discussed by 
De Ruiter et al. (\cite{Der02}); Capetti et al. (\cite{Cap02}) discussed the optical nuclei of the B2 radio galaxies 
in the context of the BLLac Unified Scheme, while Parma et al. (\cite{Par03}) presented a comprehensive search for
optical jets. 

Since some of the B2 galaxies had large amounts of dust, not all were used in the present analysis. In fact we used only 39 of the 57 B2 galaxies observed with HST, as can be seen by counting the objects listed in Tables \ref{tab:Nuker_B2dust} and \ref{tab:Nuker_B2nodust}, which are discussed in Sect. \ref{subs:fitting}. In addition to thirteen galaxies with too much dust, we excluded five other objects, either with very strong nuclei (B2 1615+32, the BL Lac B2 1101+38, and the broad line radio galaxy B2 1833+32 also known as 3C 382), or because no satisfactory one-dimensional brightness profile could be obtained (B2 0722+30 and B2 1511+26).

The HST images are a major step forward in our knowledge of the optical properties of B2 radio 
galaxies, as optical work on the B2 sample has always lagged somewhat behind the extensive
radio studies that have been done especially since the 1980's (e.g. Fanti et al. \cite{Fan87}; Parma et al. \cite{Par87}; 
De Ruiter et al. \cite{Der90}; Morganti et al. \cite{Mor97}). 
However, a complete broad band imaging survey of the B2 sample was carried out by 
Gonzalez-Serrano et al. (\cite{Gon93}) and Gonzalez-Serrano \& Carballo (\cite{Gon00}), while 
narrow band $H\alpha$ images were obtained by Morganti et al. (\cite{Mor92}). 

In order to complement the data on radio galaxies we also analyzed HST images (in the $R$ colour)
of a number of Fanaroff-Riley type II radio galaxies. 
These objects were selected from the 3CR sample and comprise all (26) 3CR radio galaxies of type FRII with
redshift $z<0.1$, similar to the range covered by the B2 sample. This is effectively
the same selection criterion used by Chiaberge et al. (\cite{Chi00}) and it leads to a starting sample of 23 objects with HST data (three not observed by HST; see Table 1 of Chiaberge et al. \cite{Chi00}). The observations were part of the HST snapshot survey of 3C radio galaxies (de Koff et al. \cite{Dek96}; Martel et al. \cite{Mar99}), while all observations were done with the $F702W$ filter (similar to the $R$ colour) and with exposure time of 280 seconds.
For a variety of reasons no reasonable Nuker fit could be obtained in eight cases: because of a complex dusty morphology, the presence of a very close companion galaxy that precluded determination of brightness profile, or incomplete coverage of the galaxy in the HST field. 
Unlike FR-I galaxies, many FR-II have bright central optical nuclei, and which led to further exclusion of five broad emission line objects and one BL Lac object (3C~371). For the latter sources the central regions of the host galaxy cannot be modelled at the smallest radii, and the parameters
describing the region inside the break in the brightness profile cannot be derived. We are thus left with a sample of nine FR~II galaxies listed in Table \ref{tab:Nuker_3C_FRII} and discussed in Sect. \ref{subs:fitting}, and for these cases we used HST archive data on which we 
performed our own analysis (including Nuker-law fits, see below). 

In this paper  we address the question of whether radio galaxies and normal ellipticals are any different in their optical
properties. In order to be able to make a proper comparison two other samples were used as well:
(i) a generic sample of mostly nearby elliptical galaxies studied by Faber et al. (\cite{Fab97}) with data obtained with the $F555W$ ($V$) filter. 
In that work the bulges of some spiral galaxies were also analysed, but for the 
sake of homogeneity we only used data from the ellipticals. And (ii) we used the sample of
Brightest Cluster Member galaxies in Laine et al. (\cite{Lai03}), observed with the $F814W$ ($I$) filter. Some basic data on the Faber et al. (\cite{Fab97}) and Laine et al. (\cite{Lai03}) samples are summarized in Table \ref{tab:FLdata}. 

We can therefore make comparisons 
between non-active (quiescent) ellipticals and "classical" radio galaxies of both the FR type I and II, 
and thus get some idea on the possible differences of the radio active and non-active galaxies.

\subsection{The surface brightness profiles and fits}
\label{subs:fitting}

One-dimensional surface brightness profiles were determined using the task ELLIPSE in the STSDAS package 
of IRAF. This program fits ellipses to the brightness of elliptical galaxies, and determines a large 
number of variables as a function of the radial distance from the galaxy-center, among which the 
most important are the brightness itself, the ellipticity and the position angle of the ellipses, and the 
parameters $a_4$ and $b_4$ (see e.g. Bender et al. \cite{Ben87b}), which are a measure of how  
"boxy" or "disky" the shapes of the isophotes are.

A bit more than one half (30/57) of the nuclear regions of B2 galaxies show conspicuous dust features in the HST images
(de Ruiter et al. \cite{Der02}); 
about three quarters of these features are in the form of circum-nuclear disks, in the other cases as more general filaments. We nevertheless 
succeeded in getting realistic profiles for 17 of the 30 dusty galaxies (cf. Table 1 of De Ruiter et al. \cite{Der02})
either by masking certain regions or by forcing certain parameters (for example the ellipticity or position angle) to remain constant in the innermost parts. 

Although theoretically the images should be deconvolved with the PSF, we decided not to do so. There are two reasons for not doing so: first the PSF of the WFPC2 is much improved with respect to the WF/PC (see Carollo et al. \cite{Car97a}; \cite{Car97b}), which makes the effects of the PSF on the brightness profiles less dramatic; second, deconvolution on our snapshot images (with moderately low signal-to-noise) produces unwanted side effects on areas with diffuse surface brightness, as is the case in the inner parts of a core profile. Trials on  several images in our sample revealed that the Lucy-Richardson method (Richardson \cite{Ric72}; Lucy \cite{Luc74}) tends to redistribute the brightness and artificially create peaks and valleys, even if the original distribution is smooth. Introduction of artefacts by the deconvolution process has been discussed by, e.g.,  Byun et al. (\cite{Byu96}).

We therefore followed Carollo et al. (\cite{Car97a}) and Carollo et al. (\cite{Car97b}), who took the similar approach of not deconvolving the direct images. Instead they convolved the one-dimensional model profiles with the PSF before doing the fit. We checked how much the Nuker parameters are affected by the PSF by creating artificial images using as various sets of Nuker parameters as input (see below) and convolving these images with a PSF created with Tinytim (Krist \cite{Kri92}). Only in cases where break radius $r_b$ is not far away from the central pixel, say $r_b < 5$ pixels, corresponding to a distance of about 0.2 arcsec, the core radius itself may have been overestimated by about 20 \% in the worst case, while the other parameters do not change significantly. Provided that $r_b$ is $> 5$ pixels, we find that variations in the Nuker parameters are negligible; in particular the inner slope $\gamma$ never varies more than 0.01. We therefore conclude that there is indeed no need for deconvolution.

In the small number of objects containing a (faint) pointlike central component we simply excluded the innermost
part from the fitting procedure, and started the fit at 0.1 arcsec outwards in 1217+29, at 0.2 arcsec in B2 0755+37, 1521+28 and 3C388, and at 
0.3 arcsec for 3C88. In all other cases the fit could be made without any restriction. In about 80 \% of the cases the fitting of the profile was done out to at least 10 arcsec, while in the remainder the profile was useful out to at least 5 arcsec (e.g. 3C 318.1).

It is common practice to fit the brightness with a "Nuker profile", an empirical law introduced by Lauer et al. (\cite{Lau95}), which has the 
advantage of being quite flexible and able to fit well especially the inner parts of the profile. In 
addition it can give a measure of the abruptness of the change of slope at the core radius. 

The Nuker law contains five parameters: core radius $r_b$ (in arcsec or pc), the brightness at
the core radius $\mu_b$ (in mag\ arcsec$^{-2}$), the outer and inner slopes of the profile ($\beta$ and $\gamma$), and
finally, parameter $\alpha$ that determines the abruptness of the change in slope around the core radius.

In terms of magnitudes the Nuker law reads:

%\begin{displaymath}
%\mu(r) = \mu_b - \left(\frac{\beta-\gamma}{\alpha}\right)2.5\log 2 + 2.5\gamma\log \frac{r}{r_b}\  +  
%2.5\left(\frac{\beta-\gamma}{\alpha}\right) \log \left(1 + (\frac{r}{r_b}) ^\alpha \right).
%\end{displaymath}

\begin{eqnarray*}
\lefteqn{\mu(r) = \mu_b - \left(\frac{\beta-\gamma}{\alpha}\right)2.5\log 2 + 2.5\gamma\log \frac{r}{r_b}\  + } \\ 
& & \hspace{3cm} + 2.5\left(\frac{\beta-\gamma}{\alpha}\right) \log \left(1 + (\frac{r}{r_b}) ^\alpha \right).
\end{eqnarray*}

It should be mentioned that according to Graham et al. (\cite{Gra03}) the Nuker law is intended to fit the innermost parts (at HST resolution 
out to $\sim 10$ arcsec or so) and was never intended to describe the outer parts of the galaxy profiles as well. According to them
a S{\'e}rsic model gives a better representation of the outer profile. Based on this
consideration Trujillo et al. (\cite{Tru04}) devised a new empirical model consisting of an outer S{\'e}rsic profile plus an inner power law. Such a model has one more free parameter than the Nuker law (six instead of five), and they claim that better fits can be obtained. Nevertheless,
in the following we stick to Nuker law profiles, as they allow easy comparison with the literature samples used here, and no S{\'e}rsic
fits are, as far as we know, available for the bulk of the galaxies described in Faber et al. (\cite{Fab97}) and Laine et al. (\cite{Lai03}).

We performed Nuker-law fits of the one-dimensional brightness profiles determined with the IRAF task ELLIPSE, using the HST images of the B2 and 3C radio galaxies.  We permitted
any values of the parameters, including negative values of $\gamma$.
The fit was done in two steps. In the first step
initial values were obtained by calculating each parameter directly from a zone were it is predominant; e.g. $\beta$ from the outer parts,
$\gamma$ in the innermost part, and $\alpha$ and $r_b$ by determining the intersection of the extrapolated inner and outer profiles, assumed to be power laws at this zero-order step. Each parameter was then varied until a minimum $\chi^2$ was reached, and this process was repeated for all other parameters. 

Of course, if we started with an arbitrary set of initial parameters,  we most likely would not converge to the correct solution (in terms of the absolute minimum of $\chi^2$), and our method would not work. So we had to start with initial values that are close to the final values. Because of the way they are determined (see above), we know that this condition is met for at least three parameters. In fact, we found that the solutions for $\beta$, $\mu_b$, and $r_b$ (and in reality in most cases also $\gamma$) were always close to the initial values, which is readily understood if we consider the nature of these parameters: $\beta$ represents the slope of outer profile and cannot be changed drastically without significantly worsening the fit. The same is true for $\mu_b$, which fixes the vertical scale and is actually the parameter that turns out to be constrained the most, and $r_b$, which sets the horizontal scale. 

\begin{figure*}
\resizebox{\hsize}{!}{
\includegraphics[width=17cm]{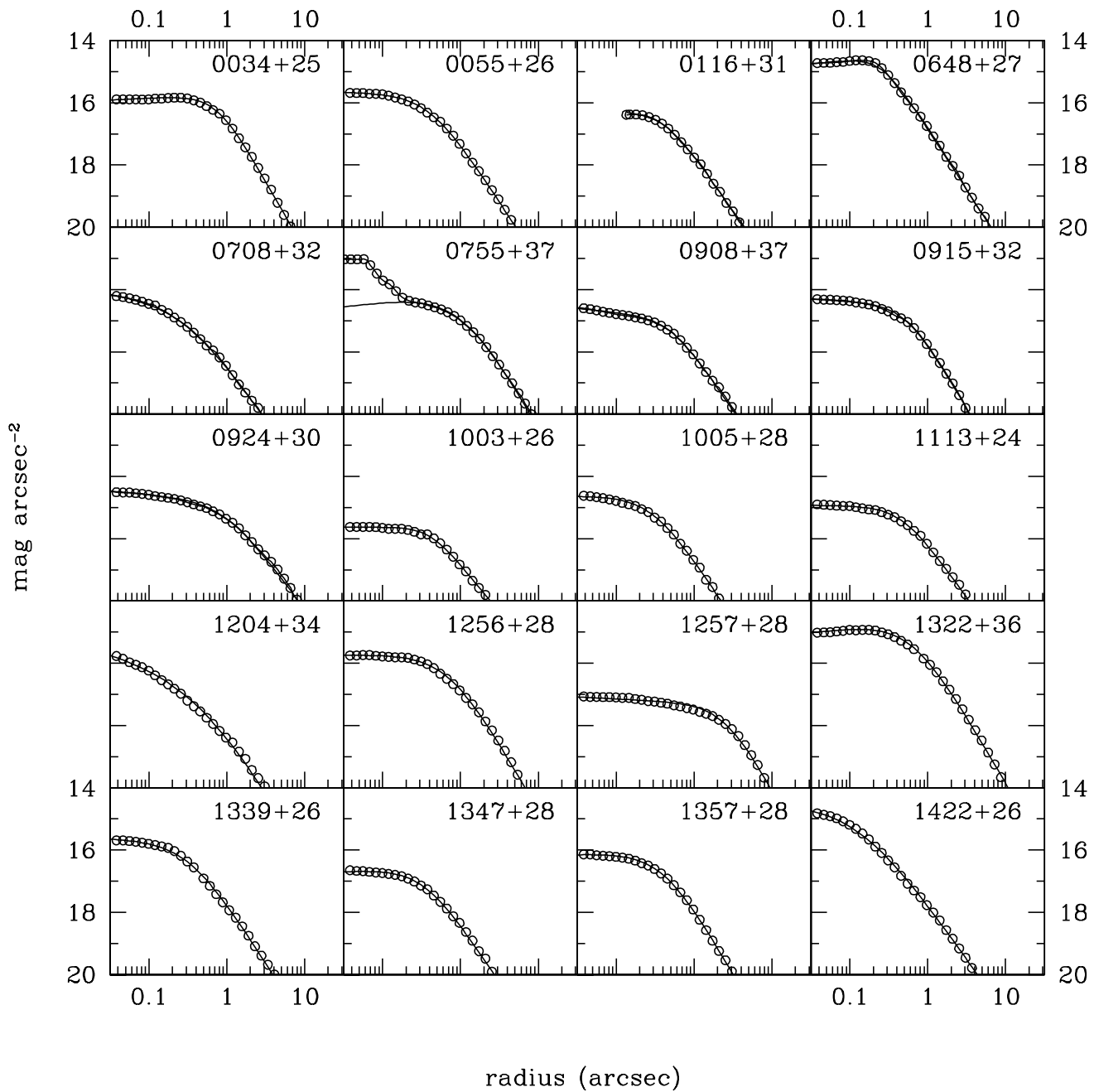}}
\caption{Radial Brightness Profiles and corresponding Nuker-law fits (see text).  The HST data of these B2 radio galaxies were part of the HST snapshot programme of B2 galaxies discussed in Capetti et al. (\cite{Cap00}). The inner region of B2 0755+37
($r<0.2$ arcsec) was excluded from the fit. }
\label{fig:fig1}
\end{figure*}

\begin{figure*}
\resizebox{\hsize}{!}{
\includegraphics[width=17cm]{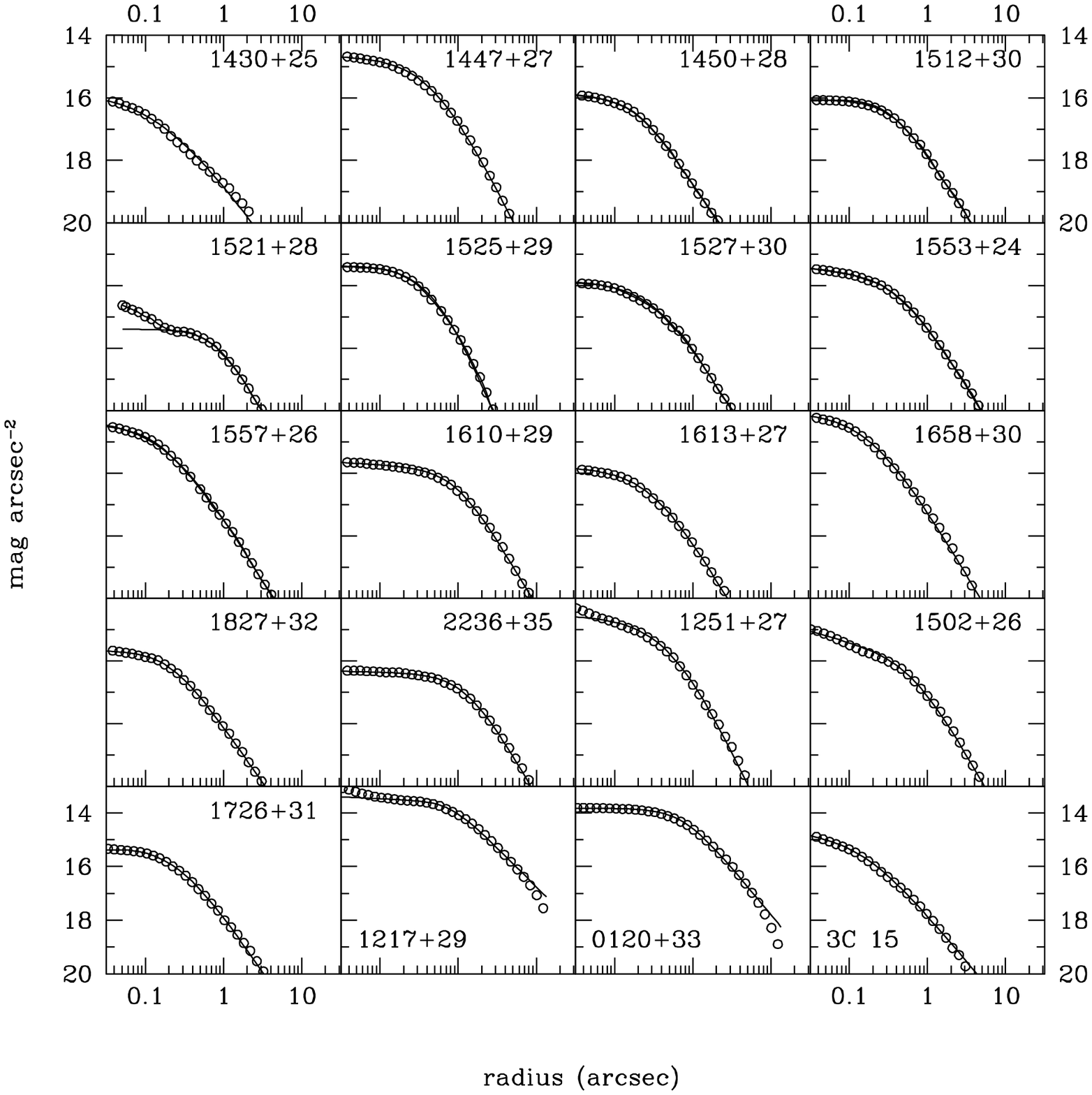}}
\caption{Radial Brightness Profiles (continued). B2 radio galaxies up to B2 2236+35 were part of the HST snapshot programme (as in Fig. \ref{fig:fig1}).  For the remaining B2 sources (from B2 1251+27 to B2 0120+33) the data were taken from the HST archive. All the B2 sources given here were discussed in Capetti et al. (\cite{Cap00}). 3C 15 is part
of the 3C FR-II sample (HST archive data). The inner region of B2 1521+28 ($r<0.2$ arcsec) was excluded from the fit.}
\label{fig:fig2}
\end{figure*}

Even so, it was possible that we would arrive only at a local $\chi^2$ minimum. Therefore we performed a second step, now using the results of the first step for $\beta$, $\mu_b$, and $r_b$ as the new initial values, and exploring a grid of initial values for $\gamma$ and $\alpha$ around their best fit values in the first step. Although the resulting Nuker parameters usually do not change significantly with respect to the first step, we sometimes did reach slightly better fits with $\gamma$ going from slightly negative to about zero, while visual inspection shows that this is indeed correct. In other cases the best fit $\gamma$ also remained (slightly) negative after the second step. Thus, the second step may in practice lead to a small improvement in the fit of the inner profile.
The errors, which are one $\sigma$ uncertainties, were determined in our Fortran programme by varying
the five Nuker law parameters step-wise, and one-by-one; they will be discussed shortly. We can now be confident that we are, within the errors, near to the absolute minimum in $\chi^2$.
The reason is that if the absolute $\chi^2$ minimum is to be found somewhere else, there must be {\it another}  one-$\sigma$ hyper-region (the region spanned by the one-$\sigma$ errors of the five parameters) centered at that point. It must correspond to a significantly different set of Nuker parameters. But we have just said that at least three of the parameters cannot be very different from the values we had already obtained, while we did verify that we arrived at a $\chi^2$ minimum by exploring the parameter space of the remaining two. Therefore the possibility that there is an "exotic" set of Nuker parameters that gives a better fit can be safely discarded.
 
Our final conclusion is therefore that our method, which has the merit of being simple and easy to use, will indeed give a final solution that corresponds to the absolute $\chi^2$ minimum, or is so close as to be indistinguishable from it, considering the errors in the parameters.

\begin{figure*}
\resizebox{\hsize}{!}{
\includegraphics[width=17cm]{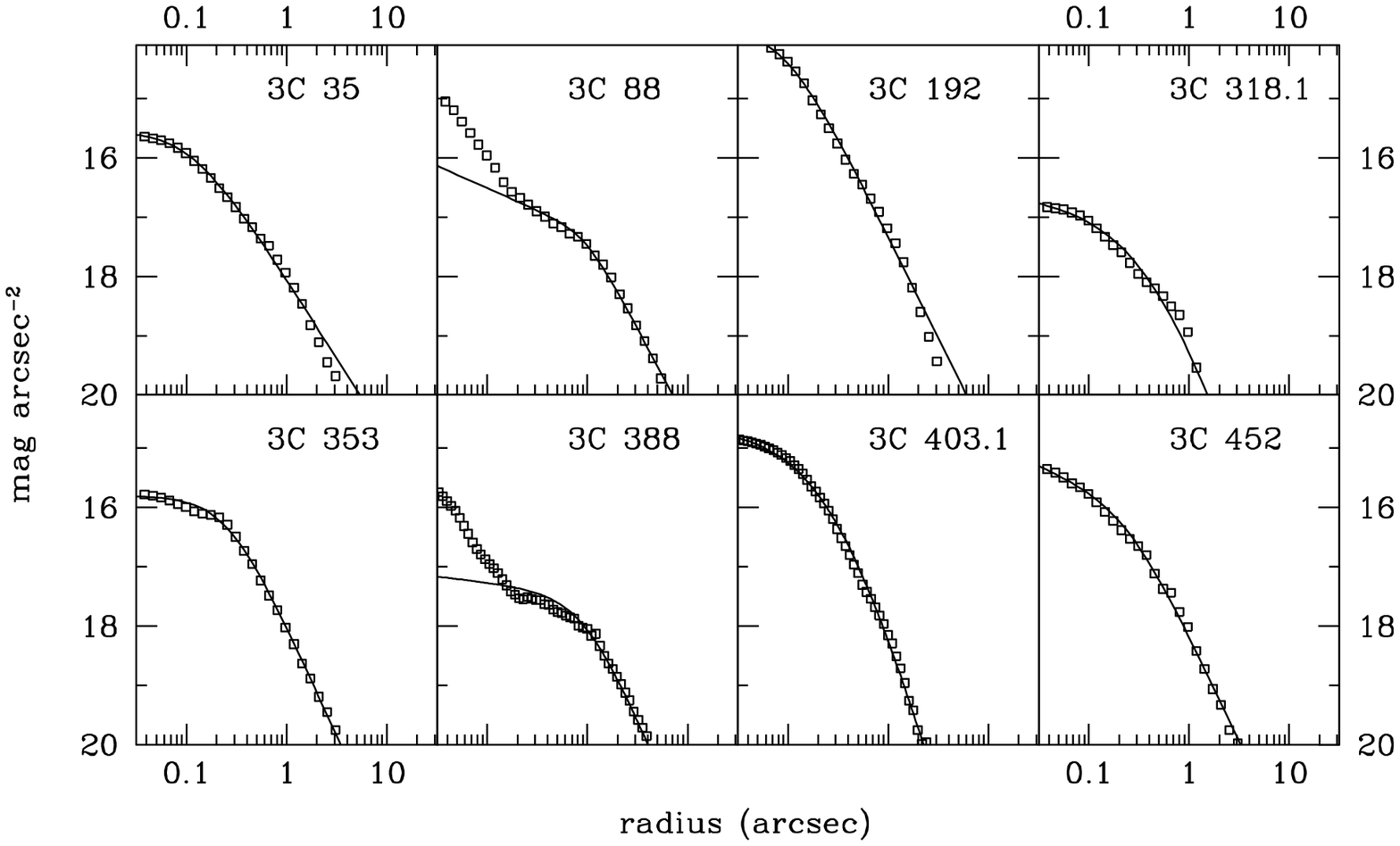}}
\caption{Radial Brightness Profiles (continued): 3C FR-II radio galaxies (HST archive data). If a pointlike nucleus is present, the inner region 
($r<0.2$ arcsec) was excluded from the fit (3C 88, 3C 388).}
\label{fig:fig3}
\end{figure*}

The results (in terms of the parameters $\alpha$, $\beta$, $\gamma$, $r_b$,
and $\mu_b$) are given in Tables \ref{tab:Nuker_B2dust}, \ref{tab:Nuker_B2nodust}, and 
\ref{tab:Nuker_3C_FRII}. Apart from the Nuker parameters we also give the radio power and absolute magnitude in $V$. In Tables 
 \ref{tab:Nuker_B2dust} and \ref{tab:Nuker_B2nodust} the radio power was taken from Capetti et al. (\cite{Cap00}), and magnitudes from the work of Gonzalez-Serrano \& Carballo (\cite{Gon00}); the magnitudes of the FR-II galaxies in Table \ref{tab:Nuker_3C_FRII} were derived mostly from
Sandage (\cite{San72}), and in a few cases from Smith \& Spinrad (\cite{Smi80}). The discreteness of this method means that, for example in the case of $\gamma$,
the one-sigma errors are often similar, since we used a course step size in the variation of the parameters; thus the error cannot be below a minimum value  (e.g. 0.05 for $\mu_b$, see the tables).
Therefore the quoted errors may in reality be
quite conservative. Of course since we try to make a fit in a five-dimensional space, the uncertainties in the derived parameter values are not mutually independent, which makes the problem of deriving error bars quite complex. The only way to proceed is by Monte Carlo analysis, which was indeed done by Byun et al. (\cite{Byu96}). They also included the possible effect of deconvolution. Since their data are quite similar to ours we may safely consider that their conclusions will also apply in our case. From their data we conclude that the true $1\sigma$ errors are in general slightly smaller than ours. The error in $\gamma$ appears to be typically 0.02 to 0.05, to be compared  with the values quoted in the tables here, which are between 0.05 to 0.15 (in the worst cases). This confirms that the errors derived in our program are realistic or even overestimated, and should in no case be underestimates.
In Tables \ref{tab:Nuker_B2dust}, \ref{tab:Nuker_B2nodust}, and \ref{tab:Nuker_3C_FRII} we also give the parameter $b_4$, the fourth cosine coefficient\footnote{There is some confusion in the literature about the name of this coefficient; often it is called $a_4$, but others use $b_4$. The important point is that boxiness is determined by the fourth cosine coefficient.} in the Fourier series expansion that measures the disk- or box-like deviations from the pure elliptical shape of an isophote; for a detailed discussion of the Fourier series expansion see Bender \& M{\"o}llenhoff (\cite{Ben87b}).
We determined the average value of $b_4$ in the interval 0.5 to 2.0 arcsec, i.e. in regions of the galaxies well outside the
nucleus, but not too far out where the background becomes comparable in strength to the galaxy brightness. The rms errors given in the tables reflect the variations within the interval in which $b_4$ was determined.

The brightness profiles of the B2, the B2/3C HST archive, and the 3C FR-II radio galaxies 
are given in Figs. \ref{fig:fig1}, \ref{fig:fig2}, and \ref{fig:fig3}. The residuals of the fitted profiles are typically $\Delta m \la 0.05$.

For the other comparison samples (nearby elliptical galaxies and brightest cluster galaxies) we used the 
values of the Nuker law parameters as published in the respective articles of Faber et al. (\cite{Fab97}) and 
Laine et al. (\cite{Lai03}) and refer to them for further information. Since the radio data are in many cases given here for the first time, we give the radio detections and upper limits based on a comparison with the NVSS in Table \ref{tab:FLdata}. The upper limits were derived from the rms-noise in the NVSS data and are $3\sigma$ limits.
For easy reference we also list the redshifts and absolute magnitudes in the same table, and make a brief comment on whether or not the object was used in our analysis. In fact, we decided to exclude dwarf galaxies with absolute magnitude greater than $-17.5$, and were not able to use some objects because no radio information is available for a variety of reasons (e.g. an object may be too far to the south to be covered by the NVSS).

There are quite a few objects with a negative inner slope ($\gamma < 0$), in particular among the B2 galaxies in which circum-nuclear
dust is present (see Table \ref{tab:Nuker_B2dust}). This tendency towards
slightly negative $\gamma$ values may be related to the presence of the dust, even if we did mask certain regions, 
and can of course be due to a small depression in the profile caused by the dust. It should be noted, however, that all negative values of $\gamma$ except one lie within one $\sigma$ of zero, so that the effect, if any, is quite small. It is relatively simple to get an estimate of the effect of dust on $\gamma$, because we have images in $V$ as well as $I$. In de Ruiter et al. (\cite{Der02}) we used this fact to get an estimate of the amount of circum-nuclear dust present in individual galaxies. Taking the galaxy B2 0034+25 as a representative case we apply a correction to the $I$ brightness in order to correct for the absorption by dust and find that $\gamma$ would be steeper by about 0.02 in the innermost parts, i.e. at a radius $<3-4$ pixels; however, {\it such regions had already been masked}. We conclude therefore that apparent flattening of the profile due to the presence of circum-nuclear dust is negligible with $\Delta\gamma<0.01$ for all galaxies that were used in our analysis.
Therefore we do not see any objects of the type 
discussed by Lauer et al. (\cite{Lau02}), which appear to have a genuine central depression in the stellar surface density.

In Table \ref{tab:nukparams} we give the median values of the Nuker-law parameters $\alpha$, $\beta$, and $\gamma$, along with the median redshifts, for all the samples used in this paper. There are no big differences in the median values of the Nuker parameters; in particular the radio galaxies, either
of the FR-I or the FR-II type, appear to be quite similar. At most the FR-I radio galaxies may have a slightly steeper outer slope $\beta$
compared to the optically selected early-type galaxies of the Laine et al. and Faber et al. samples. This may be due to the fact that these samples contain more nearby galaxies (median redshifts 0.0372 and 0.0048 for Laine et al. and Faber et al. against about 0.060 for the radio selected objects), and we may thus sample the outer slope in slightly different regions. It has been shown 
by Graham et al. (\cite{Gra03}) that the Nuker parameters are not necessarily robust against variations of the radial fitting range used. This is of course easily understandable, as fitting with five parameters requires a sufficient coverage of all parts of the profile. Moreover, the profiles tend to steepen in the outermost parts, curvature that cannot be accounted for by the Nuker parameters. This means that more distant galaxies tend to have steeper $\beta$, which may very well be reflected in the median $\beta$ given in Table \ref{tab:nukparams}. Also the other Nuker parameters will be affected, as discussed by Graham et al. (\cite{Gra03}). In particular caution is needed with the break radius $r_b$, which tends to be overestimated due to curvature in the outer profile. The effect on the inner slope $\gamma$ appears to be somewhat less severe. Of the four samples discussed presently, only the Faber et al. sample is very different in redshift and therefore in radial range covered by the brightness profile. Therefore any systematic uncertainty of the type described by Graham et al. (\cite{Gra03}) will affect the comparison between the nearby and the more distant galaxies, and may be as much as a factor 2 to 3 in the break radius. 
However, the trends in the Nuker parameters, discussed below, are so strong that they can in no way be explained by the Graham et al. selection effect alone; at most the precise form of some of the correlations involving $r_b$ (see Sect. \ref{subs:bh_fp}) is open to some doubt.

\section{Discussion}
\label{sec:discussion}

\subsection{Optical profiles and radio power}
\label{subs:optrad}

Since we want to analyze if there are any differences between radio-active and quiescent galaxies, we must know the radio
power or its upper limit.
The Laine et al. (\cite{Lai03}) and Faber et al. (\cite{Fab97}) samples were optically selected and it was
not known a priori (at least in many cases) if some galaxies of these samples are radio emitters. We searched the NVSS
1.4 GHz maps for possible radio counterparts and checked in the NED database if radio flux densities were already known.
As mentioned in the previous section the detections and upper limits are given in Table \ref{tab:FLdata}.

\begin{figure}
\resizebox{\hsize}{!}{
\includegraphics{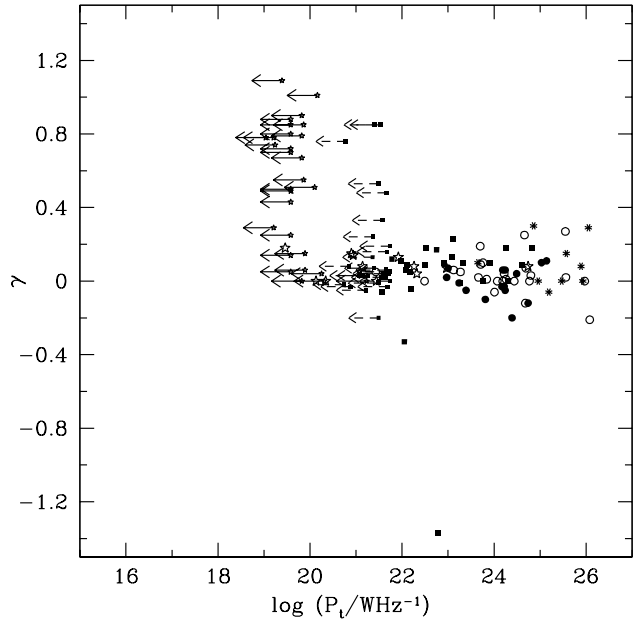}}
\caption{Log of radio power at 1.4 GHz against $\gamma$. Filled circles: B2 galaxies with dust; open circles: B2 galaxies without nuclear dust;
asterisks: 3C FR-II galaxies; filled squares: Laine et al. galaxies; starred: Faber et al. galaxies. The limits are mostly from the Faber et al.
sample (continuous arrows), with a few from the Laine et al. sample (dashed arrows).}
\label{fig:fig4}
\end{figure}

We divided the galaxies into two classes, the radio-active and the non-active 
objects. As it happens, all radio upper limits are less than $10^{21.75}$ WHz$^{-1}$ (see Fig. \ref{fig:fig4}), and only very few radio detections are below this; therefore, we can use $10^{21.75}$ WHz$^{-1}$  as the dividing line between radio-active and non-active galaxies.

Is there any difference in the profile properties that can be ascribed to a difference in radio power? 
Plotting the radio power 
against the inner slope $\gamma$ in Fig. \ref{fig:fig4} we notice a clear distinction: all radio detected objects have shallow slopes ($\gamma<0.3$), 
while the quiescent galaxies occupy the range $0<\gamma<1$ with many having steep ($>0.3$) inner slopes. 
It should also be noted that the few radio detected galaxies at low radio power ($<10^{21.75}$WHz$^{-1}$) all have a slope $\gamma < 0.3$.

It is known that weaker galaxies ($M_v > -20.5$, see Faber et al. \cite{Fab97}) tend to have power-law
profiles, and the brighter galaxies ($M_v < -22$) core profiles, which means that we do expect the fainter galaxies to
have steeper inner slopes. It might be possible that the effect shown in Fig. \ref{fig:fig4} is in reality a consequence of
the underlying difference between bright and faint galaxies.

\begin{figure}
\resizebox{\hsize}{!}{
\includegraphics{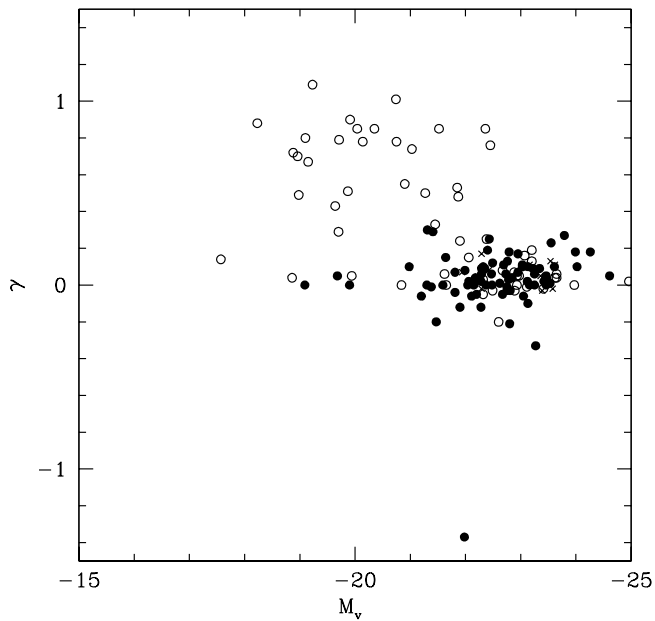}}
\caption{Absolute visual magnitude against $\gamma$. Filled points are radio-active, open circles non-active objects.
Objects with unknown radio power are indicated by crosses.}
\label{fig:fig5}
\end{figure}

Figure \ref{fig:fig5} shows that this is not the case. While it is true that the brightest galaxies ($M_v<-22.5$) are all of the core-type,
there is a considerable region of overlap in absolute magnitude (roughly $M_v<-19$ ), in which the radio-active galaxies are
of the core-type while the quiescent ones show a broad range of inner slopes, going from core- to power-law type profiles.
The only selection effect which may play some minor role is the different sampling of distances by faint and bright galaxies, which may bias the values of the Nuker-fit parameters (as discussed by Graham et al. \cite{Gra03}; see previous section).

The shape of the isophotes has been linked to other properties of the galaxies, like strong X-ray emission
(Pellegrini \cite{Pel99}; she also notes that such galaxies tend to have core-profiles), or strong radio emission:
according to  Bender et al. (\cite{Ben87a}; \cite{Ben89}), radio galaxies more often have boxy isophotes.
No shape information is available for the
Faber et al. (\cite{Fab97}) and Laine et al. (\cite{Lai03}) samples, so we can only analyze the radio-active
objects from the B2 and 3C samples. First, we note that shape parameter $b_4$, which measures the
boxiness of an isophote (see e.g. Bender et al. \cite{Ben87a}; Bender \& M{\"o}llenhoff \cite{Ben87b};
Lauer \cite{Lau85a}; Lauer \cite{Lau85b}), varies considerably from the inner to the outer isophotes (an effect
noted also by Rest et al. \cite{Res01}) and may indeed be negative in some parts and positive in other regions.
In fact, Rest et al. (\cite{Res01}) comment that there is a bewildering variety of changes in the isophotal shape, which
may oscillate from boxy to disky and vice-versa. Another problem is that Bender et al. (\cite{Ben87a}; and other
authors) measure $b_4$ in the outer parts of the galaxy, while the high resolution HST images analyzed by us cover 
the inner parts.

\begin{figure}
\resizebox{\hsize}{!}{
\includegraphics{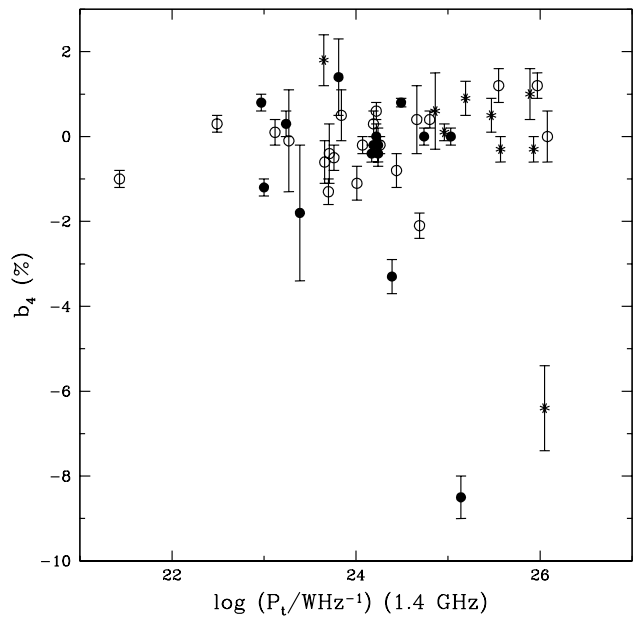}}
\caption{The logarithm of radio power against the "boxiness" coefficient $b_4$ (in \%). Filled circles: B2 radio galaxies with nuclear dust; open circles: B2 radio galaxies without nuclear dust; asterisks: 3C FR-II radio galaxies. Errorbars represent the rms variations of $b_4$ in the radial range between 0.5 and 2 arcsec.}
\label{fig:fig6}
\end{figure}
  
In Fig. \ref{fig:fig6} we show the dependence of $b_4$ on radio power. It is obvious that we do not find any significant effect.
It is true that the four objects with significant boxy isophotes ($b_4<-2 \%$) all have radio power $>10^{24}$ WHz$^{-1}$,
but the large majority of the radio-active galaxies plotted in Fig. \ref{fig:fig6} have $b_4$ close to zero. A similar plot results if
we use absolute magnitude instead of radio power. We therefore have to conclude that we cannot confirm that active galaxies
(in this case radio galaxies) tend to have boxy isophotes, but we repeat once again that this may be due to the fact that we had
to measure $b_4$ much more inward than was done in earlier works.

\subsection{The core fundamental plane and black hole masses}
\label{subs:bh_fp}

\begin{figure}
\resizebox{\hsize}{!}{
\includegraphics{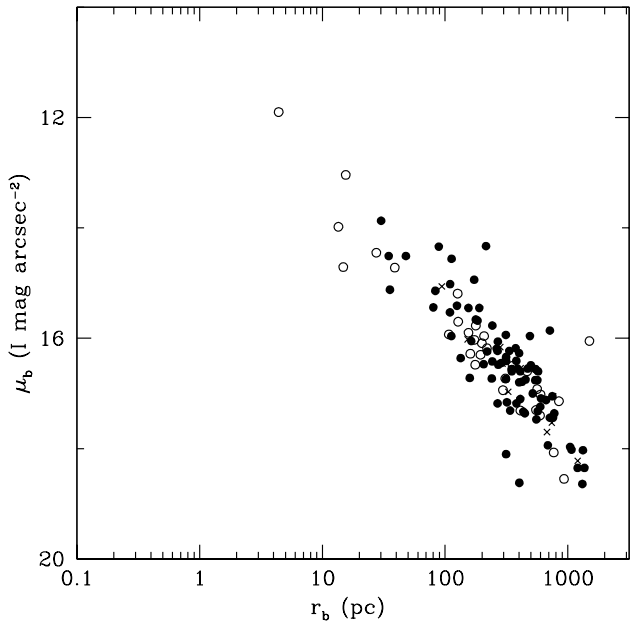}}
\caption{The core radius $r_b$ (in pc) against $\mu_b$ (in $I$ mag~arcsec$^{-2}$), for radio-active (filled circles) and non-active (open circles)
galaxies. Objects with unknown radio power are indicated by crosses. Only core-type galaxies ($\gamma < 0.3$) are shown.}
\label{fig:fig7}
\end{figure}

The core radius $r_b$ correlates tightly with the core brightness $\mu_b$, which is the equivalent of the Kormendy relation between effective scale length $r_e$ and brightness
$\mu_e$ (Kormendy \cite{Kor77}). The corresponding plot is given in Fig. \ref{fig:fig7}. We used only elliptical galaxies of the core-type
(i.e. $\gamma < 0.3$), as the interpretation of $r_b$ as a core radius is dubious in the case of power-law galaxies. 

Figure \ref{fig:fig7} confirms the correlation already found by Faber et al. (\cite{Fab97}) and sustains their conclusions 
concerning the core fundamental plane. In particular it gives us a way (admittedly rough) to estimate the distance to an object
by measuring the two observable parameters, surface brightness and core size.

\begin{figure}
\resizebox{\hsize}{!}{
\includegraphics{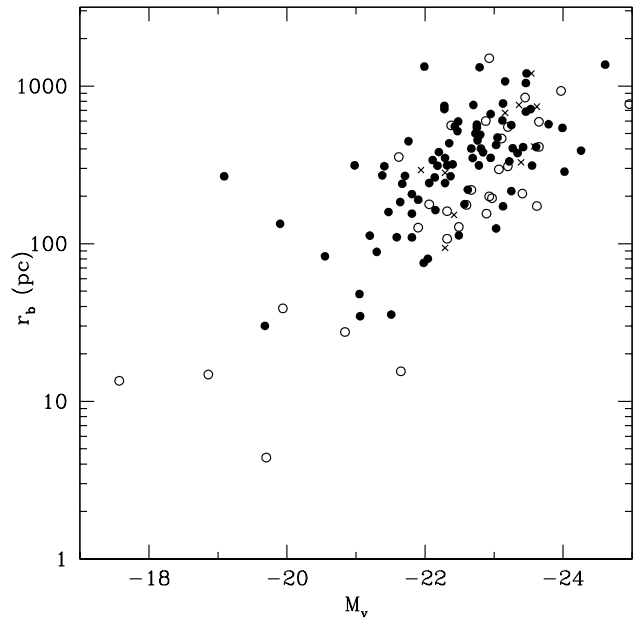}}
\caption{Absolute visual magnitude against $r_b$ for galaxies with core-type profiles. Filled circles are radio-active galaxies and open
circles non-active galaxies.
Objects with unknown radio power are indicated by crosses.}
\label{fig:fig8}
\end{figure}

In Fig. \ref{fig:fig8} we show the correlation between the absolute magnitude and the core radius. Also in this figure we used only
core-type galaxies. This relation
was discussed by Laine et al. (\cite{Lai03}), who used their own data and those of Faber et al. (\cite{Fab97}). Our best fit gives
$r_b \propto L_v^{1.05\pm 0.10}$, where $L_v$ is the optical luminosity (at the $V$ band), to be compared with
$r_b\propto L_v^{0.72}$ (Laine et al. \cite{Lai03}) and $r_b\propto L_v^{1.15}$ (Faber et al. \cite{Fab97}). 

Considering the scatter
in the data it is probably fair to say that the present data suggest a roughly linear dependence ($r_b \propto L_v$).
It should be kept in mind, however, that small break radii can only be detected in nearby galaxies. The brighter galaxies are on average at larger distances, so that one expects them to have large break radii because of this selection effect. Even if the effect is difficult to quantify, there are two reasons why it should be small. First, some of the brightest galaxies in Fig. \ref{fig:fig8} happen to be relatively nearby, and break radii of the order of 50 pc could easily have been detected. Second, distant bright galaxies with small break radii would appear as "power-law" galaxies, and we have seen in Fig. \ref{fig:fig5} that there are no such objects.

The shape of the profile may depend on the presence of massive black holes in the center, and the characteristic core-type
brightness behavior may be a signature of potential nuclear activity that is able to produce a strong radio source. From this point of view one may consider galaxies with an inner core profile as "radio active" galaxies, so that whether or not a radio source is present is then related to the particular stage in the life-cycle of the radio source, which is known to be relatively short, $< 10^8$ years (see e.g. Parma et al. \cite{Par99}).
 
There are several 
ways involving massive black holes to produce a central cusp in the density profile, for example, through the adiabatic growth of a 
single black hole (van der Marel \cite{Van99}). However, such a model would produce steeper inner profiles than the ones observed and
would not lead to the characteristic "shallow" core ($\gamma < 0.3$). Although such a model may be applicable to many types of nuclear activity found in practically all kinds of galaxies, it apparently cannot explain why radio activity occurs only in core-type galaxies.

It is more plausible to link radio activity to a rapidly spinning black hole, since the spin axis would provide a natural direction along which the radio jets would emanate, and thus at least some properties of the radio source would find their origin in the physics of the central engine. However, once again the reason that radio activity occurs in core-type galaxies remains unknown.

There is one model that gives a natural explanation for both radio activity and the creation of a shallow inner brightness profile: 
binary black hole merging
(Ravindranath et al. \cite{Rav02}; Milosavljevi{\'c} et al. \cite{Mil02}). For a detailed
discussion of the merits of single and binary black hole models we refer to Merritt (\cite{Mer04}).
The idea that the merging of two galaxies, both with massive black holes in their centers, may lead to the formation of 
a rapidly spinning and very massive central black hole, which may be the progenitor of a powerful radio source, 
was discussed by Wilson \& Colbert (\cite{Wil95}).

In short, a radio source can be present if the parent galaxy has a core-profile, which is in agreement with the idea that 
there is an innermost zone of stellar depletion, created by a (binary) black hole in the center. Of course, a core profile 
is not a sufficient condition for producing a radio source, but according to the line of reasoning given here, it should be indicative of its 
potential presence; i.e., the radio activity cycle is much shorter than the lifetime of a galaxy, so that at any given time the radio source may
be present, or may not.

\begin{figure}
\resizebox{\hsize}{!}{
\includegraphics{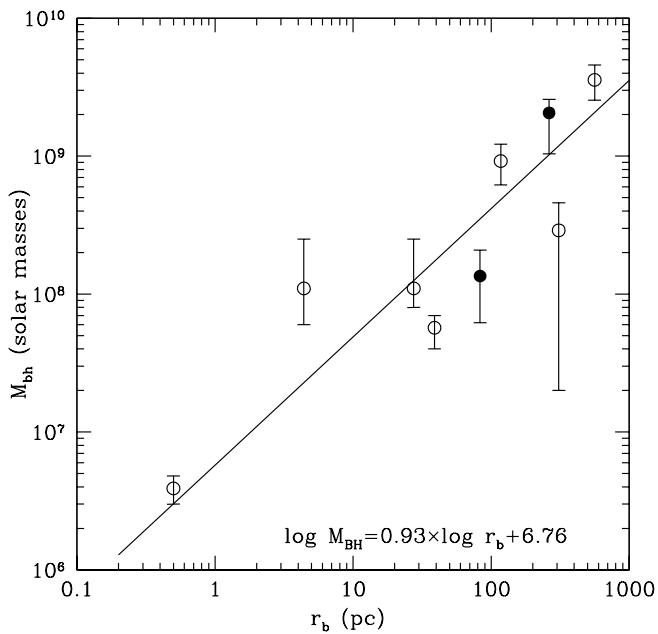}}
\caption{Core radius $r_b$ (in pc) against Black Hole mass (in solar masses) for nine galaxies of the Faber et al. sample. Filled circles
are radio-active, open circles non-active objects. Black Hole masses and their uncertainties were taken from the compilation of Merritt \& Ferrarese (\cite{Mer01}).}
\label{fig:fig9}
\end{figure}

An interesting correlation was found by Graham et al. (\cite{Gra01}). In addition to the well-known correlation between central velocity
dispersion and black hole mass, they also found that the concentration index (based on S{\'e}rsic profiles) is quite tightly linked to black hole mass, in the sense that a higher concentration of light corresponds to higher black hole mass. It is also known that the absolute bulge magnitude correlates with black hole mass (Magorrian et al. \cite{Mag98}; Wandel \cite{Wan99}; Ferrarese \& Merritt \cite{Fer00}; McLure \& Dunlop \cite{Mcl01}; Erwin et al. (\cite{Erw04}).
In the light of these usually very good correlations one might suspect that some of the Nuker law parameters may also correlate with 
the mass of the supermassive central (binary) black hole. We therefore selected a number of galaxies from the Faber et al. (\cite{Fab97}) list, for which the black hole mass has been determined (Merritt \& Ferrarese \cite{Mer01}). We indeed recover the correlation between absolute
magnitude and $M_{BH}$, but find that there is an almost equally good relation between the core radius $r_b$ and $M_{BH}$. The best fit 
gives $\log M_{BH} \propto (0.93\pm 0.19)\log r_b$ with correlation coefficient  $r^2=0.86$; it is shown in Fig. \ref{fig:fig9}. 

It is worth noting that if as the x-axis we use the variable $0.787*\log r_b - 0.256*\mu_b + 2.73$, which according to Faber et al. \cite{Fab97}, represents one of the axes of the fundamental plane,
then the correlation becomes slightly tighter (correlation coefficient $r^2=0.89$).

The fundamental plane has been studied by a number of authors, and it is now clear that some global parameters combine in such a way that
they are related to the black hole mass (see e.g. Woo et al. \cite{Woo04}; Bettoni et al. \cite{Bet01}; Bettoni et al. \cite{Bet03}).

\section{Summary and conclusions}
\label{sec:conclusions}
We have made a comparative study of several samples of early type galaxies, in order to analyze possible differences
between radio-active and non-active objects. It so happens that the four samples used can be divided into radio detected galaxies
with $P > 10^{21.75}$ WHz$^{-1}$, and upper limits: there are only a few detections below this limit in radio power.
The quiescent galaxies come from the optically selected
samples of Faber et al. (\cite{Fab97}) and Laine et al. (\cite{Lai03}), while the large majority of the radio detected galaxies
are either from the B2 sample (Capetti et al. \cite{Cap00}) or from nine 3C galaxies without a strong nuclear component, selected from HST archival data.

"Nuker-law" fits of the surface brightness profiles were carried out and Nuker parameters determined from the HST images of
the B2 and 3C samples. Nuker law parameters of the optically selected galaxies were taken from the original papers.

The results of our analysis can be summarized as follows:

\begin{itemize}
\item A clear distinction can be made between radio-active and radio quiescent galaxies. All radio-detected galaxies considered here have
shallow inner profiles with $\gamma < 0.3$, while the non-active objects (usually $P < 10^{21.75}$ WHz$^{-1}$) may be either of the
core-type or the power law type (i.e. $\gamma > 0.3$). Note that there are no differences between radio-active and radio-quiescent core-type
galaxies, and it is quite plausible that the difference in radio-activity depends on the particular stage in the life of a radio source; the
apparently quiescent galaxies may, therefore, have had a radio source in the past or will become strong radio emitters in the future once a new cycle of activity starts. A power-law galaxy should never produce a radio source. 
\item The distinction in profile between radio-active and non-active galaxies is not due to an effect with absolute magnitude, because in
the interval where the absolute magnitudes the two types of galaxies overlap the differences in the inner profile remain.
\item We do not find any differences between FR-I and FR-II radio galaxies, although it can be seen from the profiles that even the 3C galaxies
studied here (which were selected as having no strong nuclear component) nevertheless tend to have an unresolved optical core in the center.
\item We recover the well-known fundamental plane correlation between radius and corresponding surface brightness; however, as in Faber et al. (\cite{Fab97}), the correlation is between the {\it core} fundamental plane parameters break radius $r_b$ and break brightness $\mu_b$.
\item The luminosity (in our case in the $I$-band) correlates well with break radius $r_b$. We find that $r_b \propto L_v^{1.05 \pm 0.10}$,
almost exactly in between the relations found earlier by Faber et al. (\cite{Fab97}) and Laine et al. (\cite{Lai03}), so we conclude that
the break radius and luminosity are roughly proportional. 
\item Using the galaxies with independently measured black hole masses we find that there is an excellent correlation with break radius
$r_b$. The correlation becomes even a bit tighter if instead of $r_b$ we consider a fundamental plane parameter that represents the central
velocity dispersion (a combination of $r_b$ and $\mu_b$).

\end{itemize} 

\begin{acknowledgements}
This project was supported by the Italian Space Agency (ASI), under contract CNR-ASI I/R/068/02.
This research made use of the NASA/IPAC Extragalactic Database (NED) which is operated 
by the Jet Propulsion Laboratory, California Institute of Technology, under contract with the National
Aeronautics and Space Administration.
\end{acknowledgements}

\clearpage
\begin{table*}
\caption[]{Nuker law parameters of B2 radio galaxies with detected central dust}
\label{tab:Nuker_B2dust}
\begin{flushleft}
\begin{tabular}{lllrcrcccr}
\hline\noalign{\smallskip}
Name & $\alpha$ & $\beta$ & $\gamma$ & $r_b$ (arcsec) & $\mu(r_b)$ & redshift & $\log(P_t$/WHz$^{-1})$ & $M_v$ & $b_4 (\%) $ \\
          & \llap{$\pm$}$\Delta\alpha$ & \llap{$\pm$}$\Delta\beta$ &  \llap{$\pm$}$\Delta\gamma$ & 
\llap{$\pm$}$\Delta$r & \llap{$\pm$}$\Delta\mu$ & & 1.4 GHz & & $\pm\Delta b_4$  \\
\noalign{\smallskip}
\hline\noalign{\smallskip}
0034+25 & 2.6 & 1.77 & $-$0.05 & 0.74 & 16.27 & 0.0321 & 23.39 & $-$22.67 & $-$1.8 \\
               & 0.4 & 0.05 &  0.05 & 0.04 &   0.05 &             &           &            &  1.6   \\
0116+31 & 2.8 & 1.43 &  0.11 & 0.45 & 16.79 & 0.0592 & 25.14 & -23.03 & -8.5 \\
               & 0.5 & 0.05 &  0.15 & 0.02 &   0.05 &             &           &            & 0.5 \\
0648+27 & 3.0 & 1.56 & -0.10 & 0.26 & 14.94 & 0.0409 & 23.81 & -23.13 & 1.4 \\
               & 0.8 & 0.05 &  0.15 & 0.02 &   0.05 &             &           &            & 0.9 \\
0908+37 & 2.3 & 1.49 &  0.10 & 0.50 & 17.36 & 0.1040 & 25.03 & -23.13 & 0.0 \\
               & 0.8 & 0.05 &  0.05 & 0.05 &   0.10 &             &           &            & 0.2 \\
0915+32 & 2.4 & 1.70 &  0.06 & 0.52 & 17.00 & 0.0620 & 24.19 & -22.47 & -0.2 \\
               & 0.4 & 0.05 &  0.05 & 0.03 &   0.05 &             &           &            & 0.2 \\
1217+29 &  2.8   & 1.16 &  0.05 & 0.801 & 13.87 & 0.0021 & 21.43 & -19.68 & -1.0 \\
{\scriptsize (NGC 4278)} & 1.0 & 0.05 & 0.05 & 0.064 & 0.05 &    &   &     & 0.2  \\
1256+28 & 1.7 & 1.74 & -0.01 & 0.70 & 16.48 & 0.0224 & 23.24 & -21.38 & 0.3 \\
               & 0.4 & 0.15 &  0.05 & 0.07 &   0.15 &             &           &            & 0.3 \\
1322+36 & 1.7 & 1.73 & -0.12 & 0.62 & 15.45 & 0.0175 & 24.74 & -21.90 & 0.0 \\
               & 0.3 & 0.05 &  0.05 & 0.05 &   0.05 &             &           &            & 0.2 \\
1339+26 & 1.7 & 1.57 &  0.04 & 0.32 & 16.41 & 0.0757 & 24.49 & -22.84 & 0.8 \\
               & 0.6 & 0.05 &  0.15 & 0.04 &   0.15 &             &           &            & 0.1 \\
1347+28 & 1.9 & 1.51 & -0.05 & 0.34 & 17.18 & 0.0724 & 24.24 & -22.20 & -0.4 \\
               & 0.4 & 0.05 &  0.10 & 0.03 &   0.05 &             &           &            & 0.3 \\
1357+28 & 1.8 & 1.74 & -0.03 & 0.40 & 16.80 & 0.0629 & 24.22 & -22.81 & 0.0 \\
               & 0.3 & 0.05 &  0.05 & 0.03 &   0.05 &             &           &            & 0.3 \\
1430+25 & 0.8 & 1.48 & -0.20 & 0.13 & 16.72 & 0.0813 & 24.39 & -21.47 & -3.3 \\
               & 0.3 & 0.15 &  0.40 & 0.03 &   0.25 &             &           &            & 0.4 \\
1447+27 & 1.43 & 2.01 & 0.02 & 0.47 & 15.77 & 0.0306 & 22.97 & -22.06 & 0.8 \\
               & 0.25 & 0.10 & 0.15 & 0.04 &   0.15 &             &           &            & 0.2 \\
1525+29 & 1.5 & 2.56 & -0.03 & 0.53 & 16.56 & 0.0653 & 24.17 & -22.75 & -0.4 \\
               & 0.3 & 0.15 &  0.15 & 0.05 &   0.15 &             &           &            & 0.2 \\
1527+30 & 1.6 & 1.63 &  0.06 & 0.34 & 16.76 & 0.1143 & 24.24 & -23.25 & -0.2 \\
               & 0.4 & 0.05 &  0.15 & 0.03 &   0.10 &             &           &            & 0.4 \\
1557+26 & 1.26 & 1.68 & 0.07 & 0.21 & 15.45 & 0.0442 & 23.00 & -21.81 & -1.2 \\
               & 0.15 & 0.05 & 0.05 & 0.01 &   0.05 &             &           &            & 0.2 \\
1726+31 &  1.3   & 1.77 & -0.21 & 0.223 & 15.96 & 0.1670 & 26.06 & -22.80 & 0.0 \\
{\scriptsize (3C 357)}    & 0.3 & 0.10 & 0.25 & 0.027 & 0.15 &         &     &     & 0.6 \\
\noalign{\smallskip}
\hline
\end{tabular}
\end{flushleft}
\end{table*}

\newpage
\begin{table*}
\scriptsize
\caption[]{B2 radio galaxies: sources without dust}
\label{tab:Nuker_B2nodust}
\begin{flushleft}
\begin{tabular}{lllrcrcccr}
\hline\noalign{\smallskip}
Name & $\alpha$ & $\beta$ & $\gamma$ & $r_b$ (arcsec) & $\mu(r_b)$ & redshift & $\log(P_t$/WHz$^{-1})$ & $M_v$ & $b_4$ (\%) \\
          & \llap{$\pm$}$\Delta\alpha$ & \llap{$\pm$}$\Delta\beta$ &  \llap{$\pm$}$\Delta\gamma$ & 
\llap{$\pm$}$\Delta$r & \llap{$\pm$}$\Delta\mu$ & & 1.4 GHz & &  $\pm\Delta b_4$ \\
\noalign{\smallskip}
\hline\noalign{\smallskip}
0055+26 &  1.9   & 1.54 &  0.03 & 0.405 & 16.34 & 0.0472 & 24.80 & -22.78 & 0.4 \\
               &  0.3   & 0.05 &  0.05 & 0.018 &   0.05 &             &           &            & 0.2 \\
0120+33 &  2.1   & 1.41 &  0.00 & 0.746 & 15.13 & 0.0164 & 22.49 & -23.25 & 0.3 \\
{\scriptsize (NGC 507)} & 0.5 & 0.10 & 0.05 & 0.068 & 0.05 &      &       &      & 0.2 \\
0708+32 &  1.6   & 1.39 &  0.19 & 0.300 & 17.16 & 0.0672 & 23.70 & -22.40 & -1.3 \\
               &  0.3   & 0.05 &  0.05 & 0.018 &   0.05 &             &           &            & 0.3 \\
0755+37 &  1.5   & 1.80 & -0.12 & 1.092 & 17.05 & 0.0413 & 24.68 & -22.28 & - \\
               &  0.4   & 0.15 &  0.25 & 0.091 &  0.15  &             &           &            & - \\
0924+30 &  1.8   & 1.36 &  0.09 & 0.956 & 17.33 & 0.0266 & 23.71 & -22.35 & -0.4 \\
               &  0.8   & 0.15 &  0.05 & 0.136 &   0.15 &             &           &            & 0.7 \\
1003+26 &  3.0   & 1.23 &  0.00 & 0.410 & 17.94 & 0.1165 & 25.97 & -23.46 & 1.2 \\
               &  1.5   & 0.05 &  0.15 & 0.046 &   0.15 &             &           &            & 0.3 \\
1005+28 &  1.9   & 1.53 &  0.00 & 0.296 & 17.24 & 0.1476 & 24.44 & -22.48 & -0.8 \\
               &  0.4   & 0.05 &  0.05 & 0.018 &   0.05 &             &           &            & 0.4 \\
1113+24 &  2.2   & 1.42 &  0.01 & 0.473 & 17.44 & 0.1021 & 23.84 & -23.53 & 0.5 \\
               &  0.4   & 0.05 &  0.05 & 0.027 &   0.05 &             &           &            & 0.6 \\
1204+34 &  1.0   & 1.72 &  0.25 & 0.455 & 17.47 & 0.0788 & 24.66 & -22.43 & 0.4 \\
               &  0.6   & 0.25 &  0.25 & 0.091 &   0.25 &             &           &            & 0.8 \\
1251+27 &  1.3   & 2.18 &  0.02 & 0.546 & 15.86 & 0.0857 & 25.56 & -22.28 & - \\
{\scriptsize (3C 277.3)} & 0.3 & 0.25 & 0.25 & 0.077 & 0.25 &          &      &   & - \\
1257+28 &  2.1 & 1.83  & 0.05 & 2.548 & 17.97 & 0.0239 & 23.27 & -23.46 & -0.1 \\
               &  0.5 & 0.25  & 0.10 & 0.001 &   0.10 &             &           &            & 1.2 \\
1422+26 &  1.14 & 1.44 &  0.00 & 0.130 & 15.44 & 0.0370 & 24.19 & -22.04 & 0.3 \\
               &  0.15 & 0.05 &  0.15 & 0.007 &   0.05 &             &           &            & 0.3 \\
1450+28 &  2.1   & 1.52 &  0.07 & 0.195 & 16.56 & 0.1265 & 24.69 & -22.95 & -2.1 \\
               &  0.3   & 0.05 &  0.05 & 0.009 &   0.05 &             &           &            & 0.3 \\
1502+26 &  1.7   & 1.77 &  0.27 & 0.655 & 16.60 & 0.0540 & 25.55 & -23.79 & 1.2 \\
{\scriptsize (3C 310)}    & 0.6 & 0.15 & 0.05 & 0.068 & 0.15 &      &       &      & 0.4 \\
1512+30 &  2.0   & 1.57 & -0.06 & 0.337 & 16.55 & 0.0931 & 24.01 & -23.05 & -1.1 \\
               &  0.5   & 0.05 &  0.05 & 0.027 &   0.05 &             &           &            & 0.4 \\
1521+28 &  2.2   & 1.83 &  0.00 & 0.846 & 18.02 & 0.0825 & 24.77 & -23.16 & - \\
               &  0.6   & 0.05 &  0.25 & 0.059 &   0.15 &             &           &            & - \\
1553+24 &  1.7   & 1.64 &  0.10 & 0.446 & 16.41 & 0.0426 & 23.76 & -22.32 & -0.5 \\
               &  0.3   & 0.05 &  0.05 & 0.027 &   0.05 &             &           &            & 0.3 \\
1610+29 &  2.0   & 1.70 &  0.06 & 0.946 & 16.49 & 0.0313 & 23.12 & -22.73 & 0.1 \\
               &  0.3   & 0.05 &  0.05 & 0.064 &   0.05 &             &           &            & 0.3 \\
1613+27 &  1.5   & 1.69 &  0.04 & 0.305 & 16.74 & 0.0647 & 24.22 & -22.18 & 0.6 \\
               &  0.4   & 0.10 &  0.15 & 0.032 &   0.15 &             &           &            & 0.2 \\
1658+30 &  1.3   & 1.69 &  0.00 & 0.187 & 15.02 & 0.0351 & 24.07 & -21.59 & -0.2 \\
               &  0.3   & 0.05 &  0.15 & 0.018 &   0.15 &             &           &            & 0.2 \\
1827+32 &  1.91 & 1.43 &  0.01 & 0.211 & 16.24 & 0.0659 & 24.26 & -22.62 & -0.2 \\
               &  0.25 & 0.05 &  0.05 & 0.007 &   0.05 &             &           &            & 0.2 \\
2236+35 &  1.9   & 1.74 &  0.02 & 1.283 & 17.09 & 0.0277 & 23.66 & -23.12 & -0.6 \\
               &  0.4   & 0.15 &  0.05 & 0.109 &   0.05 &             &           &            & 0.5 \\
\noalign{\smallskip}
\hline
\end{tabular}
\end{flushleft}
\end{table*}

\clearpage
\begin{table*}
\caption[]{3C radio galaxies: FR-II sources}
\label{tab:Nuker_3C_FRII}
\begin{flushleft}
\begin{tabular}{lllrcrcccr}
\hline\noalign{\smallskip}
Name & $\alpha$ & $\beta$ & $\gamma$ & $r_b$ (arcsec) & $\mu(r_b)$ & redshift & $\log(P_t$/WHz$^{-1})$ & $M_v$ & $b_4$ (\%) \\
          & \llap{$\pm$}$\Delta\alpha$ & \llap{$\pm$}$\Delta\beta$ &  \llap{$\pm$}$\Delta\gamma$ & 
\llap{$\pm$}$\Delta$r & \llap{$\pm$}$\Delta\mu$ & & 1.4 GHz &  & $\pm\Delta b_4$\\
\noalign{\smallskip}
\hline\noalign{\smallskip}
3C 015    & 1.4 & 1.39 &  0.15 & 0.159 & 15.68 & 0.0730 & 25.57 & -21.64 & -0.3 \\
               & 0.4 & 0.05 &  0.15 & 0.018 &   0.10 &             &           &            & 0.3 \\
3C 035    & 1.6 & 1.07 & -0.06 & 0.105 & 15.96 & 0.0670 & 25.19 & -21.20 & 0.9 \\
               & 0.6 & 0.05 &  0.25 & 0.014 &   0.15 &             &           &            & 0.4 \\
3C 088    & 3.4 & 1.28 &  0.30 & 0.978 & 17.47 & 0.0300 & 24.86 & -21.31 & 0.6 \\
               & 2.4 & 0.15 &  0.15 & 0.086 &   0.05 &             &            &           & 0.9 \\
3C 192    & 1.8 & 1.37 &  0.00 & 0.091 & 14.34 & 0.0600 & 25.47 & -21.30 & 0.5 \\
               & 0.9 & 0.15 &  0.40 & 0.009 &   0.15 &             &           &            & 0.4 \\
3C 318.1 & 1.2 & 1.90 &  0.10 & 0.410 & 18.10 & 0.0460 & 23.65 & -20.98 & 1.8 \\
               & 0.8 & 0.60 &  0.20 & 0.136 &   0.25 &             &           &            & 0.6 \\
3C 353    & 2.0 & 1.49 &  0.00 & 0.259 & 16.36 & 0.0300 & 25.93 & -19.90 & -0.3 \\
               & 0.4 & 0.05 &  0.15 & 0.018 &   0.05 &             &           &            & 0.3 \\
3C 388    & 2.0 & 1.60 &  0.08 & 0.956 & 18.03 & 0.0910 & 25.89 & -21.99 & 1.0 \\
               & 1.0 & 0.15 &  0.25 & 0.091 &   0.15 &              &          &             & 0.6 \\
3C 403.1 & 1.1 & 2.33 &  0.00 & 0.296 & 16.23 & 0.0554 & 24.96 & -19.09 & 0.1 \\
               & 0.3 & 0.15 &  0.15 & 0.027 &   0.15 &             &           &            & 0.2 \\
3C 452    & 1.8 & 1.45 &  0.29 & 0.246 & 16.41 & 0.0810 & 26.05 & -21.41 & -6.4 \\
               & 0.9 & 0.15 &  0.15 & 0.027 &   0.15 &             &            &           & 1.0 \\
\noalign{\smallskip}
\hline
\end{tabular}
\end{flushleft}
\end{table*}

\clearpage
\begin{table*}
\caption[]{Faber et al. and Laine et al. samples: basic data}
\label{tab:FLdata}
\begin{flushleft}
\begin{tabular}{lccccl}
\hline\noalign{\smallskip}
Name & redshift & $M_v$ & $\log(P_t$/WHz$^{-1})$ & Sample & Comments \\
          &               &             & 1.4 GHz \\
\noalign{\smallskip}
\hline\noalign{\smallskip}
A1020            & 0.0650 & $-22.29$ & $-$                 & F & no radio info \\
A1831            & 0.0749 & $-23.16$ & $-$                 & F & no radio info \\
NGC  221       & 0.0002 & $-16.60$ & \llap{$<$}16.95 & F & not used \\
NGC  221V     & 0.0051 & $-16.60$ & $-$                 & F & not used \\
NGC  524       & 0.0062 & $-21.51$ & 20.34             & F \\
NGC  596       & 0.0057 & $-20.90$ & \llap{$<$}19.86 & F \\
NGC  720       & 0.0060 & $-21.62$ & \llap{$<$}19.89 & F \\ 
NGC 1023      & 0.0027 & $-20.14$ & \llap{$<$}19.21 & F \\
NGC 1172      & 0.0080 & $-20.74$ & \llap{$<$}20.16 & F \\
NGC 1399      & 0.0048 & $-21.71$ & 22.97             & F \\ 
NGC 1400      & 0.0057 & $-21.06$ & 20.13             & F \\
NGC 1426      & 0.0057 & $-20.35$ & \llap{$<$}19.86 & F \\
NGC 1600      & 0.0134 & $-22.70$ & 22.26             & F \\
NGC 1700      & 0.0095 & $-21.65$ & \llap{$<$}20.31 & F \\
NGC 2636      & 0.0089 & $-18.86$ & \llap{$<$}20.25 & F \\
NGC 3115      & 0.0022 & $-20.75$ & \llap{$<$}19.04 & F \\
NGC 3377      & 0.0027 & $-19.70$ & \llap{$<$}19.21 & F \\
NGC 3379      & 0.0027 & $-20.55$ & 19.46             & F \\
NGC 3599      & 0.0054 & $-19.71$ & \llap{$<$}19.82 & F \\
NGC 3605      & 0.0054 & $-19.15$ & \llap{$<$}19.82 & F \\
NGC 3608      & 0.0054 & $-20.84$ & \llap{$<$}19.82 & F \\
NGC 4168      & 0.0097 & $-21.76$ & 20.94             & F \\
NGC 4365      & 0.0059 & $-22.06$ & \llap{$<$}19.89 & F \\
NGC 4387      & 0.0041 & $-18.88$ & \llap{$<$}19.58 & F \\
NGC 4434      & 0.0041 & $-18.96$ & \llap{$<$}19.58 & F \\
NGC 4458      & 0.0041 & $-18.98$ & \llap{$<$}19.58 & F \\
NGC 4464      & 0.0041 & $-18.23$ & \llap{$<$}19.58 & F \\
NGC 4467      & 0.0041 & $-17.04$ & \llap{$<$}19.58 & F & not used \\
NGC 4472      & 0.0041 & $-22.57$ & 22.32             & F \\
NGC 4478      & 0.0041 & $-19.64$ & \llap{$<$}19.58 & F \\
NGC 4486      & 0.0041 & $-22.38$ & \llap{$<$}19.58 & F \\
NGC 4486B   & 0.0041 & $-17.57$ & \llap{$<$}19.58 & F \\
NGC 4551      & 0.0041 & $-19.10$ & \llap{$<$}19.58 & F \\
\noalign{\smallskip}
\hline
\end{tabular}
\smallskip
\end{flushleft}
\end{table*}

\clearpage
\begin{table*}
\addtocounter{table}{-1}
\caption[]{Faber et al. and Laine et al. samples: basic data (continued)}
%\label{tab:FLdata}
\begin{flushleft}
\begin{tabular}{lccccl}
\hline\noalign{\smallskip}
Name & redshift & $M_v$ & $\log(P_t$/WHz$^{-1})$ & Sample & Comments \\
           &              &             & 1.4 GHz \\
\noalign{\smallskip}
\hline\noalign{\smallskip}
NGC 4552      & 0.0041 & $-21.05$ & 21.44            & F \\
NGC 4564      & 0.0041 & $-19.94$ & \llap{$<$}19.58 & F \\
NGC 4570      & 0.0041 & $-20.04$ & \llap{$<$}19.58 & F \\
NGC 4621      & 0.0041 & $-21.27$ & \llap{$<$}19.58 & F \\
NGC 4636      & 0.0041 & $-21.67$ & 21.92            & F \\
NGC 4649      & 0.0041 & $-22.14$ & 20.90            & F \\
NGC 4697      & 0.0028 & $-21.03$ & \llap{$<$}19.24 & F \\
NGC 4742      & 0.0033 & $-19.23$ & \llap{$<$}19.39 & F \\
NGC 4874      & 0.0249 & $-23.54$ & $-$                & F & no radio info \\
NGC 4889      & 0.0249 & $-23.36$ & $-$                & F & no radio info \\
NGC 5813      & 0.0075 & $-21.81$ & 21.14            & F \\
NGC 5845      & 0.0075 & $-19.87$ & \llap{$<$}20.10 & F \\
NGC 6166      & 0.0300 & $-23.47$ & \llap{$<$}24.74 & F \\
NGC 7332      & 0.0054 & $-19.91$ & \llap{$<$}19.82 & F \\
NGC 7768      & 0.0275 & $-22.93$ & \llap{$<$}21.23 & F \\
VCC 1199       & 0.0041 & $-15.24$ & \llap{$<$}19.58 & F & not used \\
VCC 1545       & 0.0041 & $-17.15$ & \llap{$<$}19.58 & F & not used \\
VCC 1627       & 0.0041 & $-16.08$ & \llap{$<$}19.58 & F & not used \\
Abell   76        & 0.0372 & $-22.60$ & \llap{$<$}21.49 & L \\
Abell  119       & 0.0445 & $-23.65$ & \llap{$<$}21.65 & L \\
Abell  168       & 0.0448 & $-22.77$ & 22.13            & L & not used \\
Abell  189       & 0.0330 & $-21.52$ & \llap{$<$}21.39 & L \\
Abell  193       & 0.0485 & $-23.55$ & 23.10            & L \\
Abell  194       & 0.0179 & $-22.67$ & \llap{$<$}20.85 & L \\
Abell  195       & 0.0427 & $-22.29$ & 24.61            & L \\
Abell  260       & 0.0367 & $-23.11$ & \llap{$<$}21.48 & L \\
Abell  261       & 0.0164 & $-22.45$ & \llap{$<$}20.78 & L \\
Abell  295       & 0.0430 & $-22.76$ & 23.09            & L \\
Abell  347       & 0.0187 & $-22.49$ & \llap{$<$}20.89 & L \\
Abell  376       & 0.0490 & $-23.20$ & \llap{$<$}21.73 & L \\
Abell  397       & 0.0333 & $-22.88$ & \llap{$<$}21.40 & L \\
Abell  419       & 0.0409 & $-21.45$ & \llap{$<$}21.57 & L \\
Abell  496       & 0.0330 & $-23.61$ & 23.33            & L \\
\noalign{\smallskip}
\hline
\end{tabular}
\smallskip
\end{flushleft}
\end{table*}

\clearpage
\begin{table*}
\addtocounter{table}{-1}
\caption[]{Faber et al. and Laine et al. samples: basic data (continued)}
%\label{tab:FLdata}
\begin{flushleft}
\begin{tabular}{lccccl}
\hline\noalign{\smallskip}
Name & redshift & $M_v$ & $\log(P_t$/WHz$^{-1})$ & Sample & Comments \\
          &               &             & 1.4 GHz \\
\noalign{\smallskip}
\hline\noalign{\smallskip}
Abell  533       & 0.0472 & $-22.29$ & 22.09            & L \\
Abell  548       & 0.0398 & $-22.37$ & 23.75            & L \\
Abell  634       & 0.0271 & $-22.32$ & \llap{$<$}21.22 & L \\
Abell  779       & 0.0227 & $-23.42$ & 21.63            & L \\
Abell  912       & 0.0452 & $-21.87$ & \llap{$<$}21.66 & L \\
Abell  999       & 0.0320 & $-22.11$ & 21.57            & L \\
Abell 1016      & 0.0322 & $-21.90$ & \llap{$<$}21.37 & L \\
Abell 1142      & 0.0350 & $-22.49$ & 21.78            & L \\
Abell 1177      & 0.0318 & $-23.20$ & \llap{$<$}21.36 & L \\
Abell 1228      & 0.0366 & $-21.85$ & \llap{$<$}21.48 & L \\
Abell 1314      & 0.0328 & $-22.95$ & 22.75            & L \\
Abell 1367      & 0.0216 & $-22.69$ & 21.97            & L \\
Abell 1631      & 0.0468 & $-22.89$ & \llap{$<$}21.69 & L \\
Abell 1656      & 0.0232 & $-23.45$ & \llap{$<$}21.08 & L \\
Abell 1983      & 0.0454 & $-21.98$ & 22.78            & L \\
Abell 2040      & 0.0457 & $-23.07$ & \llap{$<$}21.67 & L \\
Abell 2147      & 0.0350 & $-22.79$ & 22.51            & L \\
Abell 2162      & 0.0321 & $-22.75$ & 23.26            & L \\
Abell 2197      & 0.0301 & $-23.27$ & 22.05            & L \\
Abell 2247      & 0.0388 & $-22.36$ & \llap{$<$}21.53 & L \\
Abell 2572      & 0.0414 & $-23.03$ & 22.90            & L \\
Abell 2589      & 0.0413 & $-23.62$ & \llap{$<$}21.58 & L \\
Abell 2877      & 0.0244 & $-23.58$ & $-$                & L & no radio info \\
Abell 3144      & 0.0452 & $-21.94$ & $-$                & L & no radio info \\
Abell 3193      & 0.0345 & $-22.42$ & $-$                & L & no radio info \\
Abell 3376      & 0.0464 & $-22.93$ & \llap{$<$}21.68 & L \\
Abell 3395      & 0.0490 & $-23.76$ & $-$                & L & no radio info \\
Abell 3526      & 0.0115 & $-24.02$ & 23.92            & L \\
Abell 3528      & 0.0544 & $-23.99$ & 24.26            & L \\
Abell 3532      & 0.0555 & $-24.26$ & 24.82            & L \\
Abell 3554      & 0.0478 & $-23.65$ & \llap{$<$}21.71 & L \\
Abell 3556      & 0.0483 & $-23.34$ & 22.50            & L \\
Abell 3558      & 0.0477 & $-24.61$ & 22.17            & L \\
\noalign{\smallskip}
\hline
\end{tabular}
\smallskip
\end{flushleft}
\end{table*}

\clearpage
\begin{table*}
\addtocounter{table}{-1}
\caption[]{Faber et al. and Laine et al. samples: basic data (continued)}
%\label{tab:FLdata}
\begin{flushleft}
\begin{tabular}{lccccl}
\hline\noalign{\smallskip}
Name & redshift & $M_v$ & $\log(P_t$/WHz$^{-1})$ & Sample & Comments \\
          &               &             & 1.4 GHz \\
\noalign{\smallskip}
\hline\noalign{\smallskip}
Abell 3562      & 0.0490 & $-23.97$ & \llap{$<$}21.73 & L \\
Abell 3564      & 0.0491 & $-22.97$ & \llap{$<$}21.73 & L \\
Abell 3570      & 0.0372 & $-22.15$ & 24.30            & L \\
Abell 3571      & 0.0397 & $-24.96$ & \llap{$<$}21.55 & L \\
Abell 3574      & 0.0155 & $-23.41$ & \llap{$<$}20.73 & L \\
Abell 3656      & 0.0192 & $-23.22$ & 22.12            & L \\
Abell 3677      & 0.0460 & $-21.81$ & 22.19            & L \\
Abell 3716      & 0.0448 & $-23.39$ & $-$                & L & no radio info \\
Abell 3736      & 0.0487 & $-23.62$ & $-$                & L & no radio info \\
Abell 3742      & 0.0161 & $-21.84$ & $-$                & L & no radio info \\
Abell 3747      & 0.0306 & $-22.29$ & $-$                & L & no radio info \\
Abell 4038      & 0.0283 & $-22.32$ & \llap{$<$}21.25 & L \\
\noalign{\smallskip}
\hline
\end{tabular}
\smallskip

N.B. F: Faber et al. (\cite{Fab97}); L: Laine et al. (\cite{Lai03}).
Detection limit (3 $\sigma$) is 1.35 mJy; some objects could not be searched
(e.g. because too far to the south, or because a precise position is lacking). Galaxies with $M_v>-17.5$ were excluded from the analysis. Abell 168 was excluded because there is a large discrepancy between the fit parameter $\gamma$ and the directly measured $\gamma_{0.05}$, i.e. the slope at 0.05 arcsec. We considered the fit parameters in contradiction with the actually observed inner profile.

\end{flushleft}
\end{table*}

\clearpage
\begin{table}
\caption[]{Median values of Nuker law parameters.}
\label{tab:nukparams}
\begin{flushleft}
\begin{tabular}{llrrr}
\hline\noalign{\smallskip}
Sample & median $z$ & \multicolumn{1}{c}{$\alpha$} & \multicolumn{1}{c}{$\beta$} & \multicolumn{1}{c}{$\gamma$}  \\
\noalign{\smallskip}
\hline\noalign{\smallskip}
B2 with dust & 0.0620 & $1.7\pm 0.2$ & $1.68\pm 0.10$ & $-0.01\pm 0.03$ \\
B2 no dust    & 0.0594 & $1.9\pm 0.2$ & $1.60\pm 0.08$ & $0.02\pm 0.02$ \\
3C FR-II         & 0.0600 & $1.8\pm 0.2$ & $1.45\pm 0.08$ & $0.08\pm 0.06$ \\
Laine et al.   & 0.0048 & $2.1\pm 0.2$ & $1.32\pm 0.02$ & $0.05\pm 0.02$ \\
Faber et al.   & 0.0372 & $2.0\pm 0.2$ & $1.33\pm 0.02$ & $0.08\pm 0.05$ \\
\noalign{\smallskip}
\hline
\end{tabular}

\smallskip
N.B. Faber et al. (\cite{Fab97}) and Laine et al. (\cite{Lai03}) samples restricted to core-type
galaxies.
\end{flushleft}
\end{table}


\begin{thebibliography}{}
\bibitem[1987a]{Ben87a}
Bender, R., D{\"o}bereiner, S., \& M{\"o}llenhoff, C. 1987a, A\&A, 177, L53
\bibitem[1987b]{Ben87b}
Bender, R., \& M{\"o}llenhoff, C. 1987b, A\&A, 177, 71
\bibitem[1989]{Ben89}
Bender, R., Surma, P., D{\"o}bereiner, S., M{\"o}llenhoff, C., \& Madejsky, R. 1989, A\&A, 217, 35 
\bibitem[2001]{Bet01}
Bettoni, D., Falomo, R., Fasano, G., et al. 2001, A\&A, 380, 471
\bibitem[2003]{Bet03}
Bettoni, D., Falomo, R., Fasano, G., \& Govoni, F. 2003, A\&A, 399, 869
\bibitem[1996]{Byu96}
Byun, Y.-I., Grillmair, C.J., Faber, S.M., et al. 1996, AJ, 111, 1889
\bibitem[2000]{Cap00}
Capetti, A., de Ruiter, H.R., Fanti, R., et al. 2000,  A\&A, 362, 871
\bibitem[2002]{Cap02}
Capetti, A., Celotti, A., Chiaberge, M., et al. 2002 A\&A, 383, 104
\bibitem[1997a]{Car97a}
Carollo, C.M., Franx, M., Illingworth, G.D., \& Forbes, D.A. 1997a, ApJ, 481, 710
\bibitem[1997b]{Car97b}
Carollo, C.M., Danziger, I.J., Rich, R.M., \& Chen, X. 1997b, ApJ, 491, 545
\bibitem[2000]{Chi00}
Chiaberge, M., Capetti, A., \& Celotti, A. 2000, A\&A 355, 873
\bibitem[1993]{Cra93}
Crane, P., Stiavelli, M., King, I. R., et al. 1993, AJ, 106, 1371
\bibitem[1996]{Dek96}
De Koff, S., Baum, S.A., Sparks, W.B., et al. 1996, ApJS 107, 621
\bibitem[1990]{Der90}
de Ruiter, H.R., Parma, P., Fanti, C. \& Fanti, R. 1990, A\&A, 227, 351
\bibitem[2002]{Der02}
de Ruiter, H.R., Parma, P., Capetti, A., Fanti, R., \& Morganti, R. 2002, A\&A, 396, 857
\bibitem[1987]{Djo87}
Djorgovski, S., \& Davis, M. 1987, ApJ, 313, 59
\bibitem[1987]{Dre87}
Dressler, A., Lynden-Bell, D., Burstein, D., et al. 1987, ApJ, 313, 42
\bibitem[2004]{Erw04}
Erwin, P., Graham, A.W., \& Caon, N. 2004, Coevolution of black holes and galaxies, Carnegie Observatories Centennial Symposia,
Carnegie Observatories Astrophysics Series, Ed. L.C. Ho
\bibitem[1997]{Fab97}
Faber, S.M., Tremaine, S., Ajhar, E.A., et al. 1997, AJ, 114, 1771
\bibitem[1987]{Fan87}
Fanti, C., Fanti, R., De Ruiter, H.R., \& Parma, P. 1987, A\&AS, 69, 57
\bibitem[2000]{Fer00}
Ferrarese, L., \& Merritt, D. 2000, ApJ, 539, L9
\bibitem[2000]{Geb00}
Gebhardt, K., Bender, R., Bower, G., et al. 2000, ApJ, 539, L13
\bibitem[1993]{Gon93}
Gonzalez-Serrano, J.I., Carballo, R., \& Perez-Fournon, I. 1993,  AJ, 105, 1710
\bibitem[2000]{Gon00}
Gonzalez-Serrano, J.I., \& Carballo, R. 2000, A\&AS, 142, 353
\bibitem[2001]{Gra01}
Graham, A.W., Erwin, P., Caon, N., \& Trujillo, I. 2001, ApJ, 563, L11
\bibitem[2003]{Gra03}
Graham, A.W., Erwin, P., Trujillo, I., \& Asensio Ramos, A. 2003, AJ, 125, 2951 
\bibitem[1986]{Hec86}
Heckman, T.M., Smith, E.P., Baum, et al. 1986, ApJ, 311, 526
\bibitem[1994]{Jaf94}
Jaffe, W., Ford, H.C., O'Connell, R.W., van den Bosch, F.C., \& Ferrarese, L. 1994, AJ, 108, 1567
\bibitem[1977]{Kor77}
Kormendy, J.G. 1977, ApJ, 217, 406
\bibitem[1992]{Kri92}
Krist, J. 1992, Tinytim v2.1 User's Manual (STScI)
\bibitem[2003]{Lai03}
Laine, S., van der Marel, R.P., Lauer, T.R., et al. 2003, AJ, 125, 478
\bibitem[1985a]{Lau85a}
Lauer, T.R. 1985a, ApJS, 57, 473
\bibitem[1985b]{Lau85b}
Lauer, T.R. 1985b, MNRAS, 216, 429
\bibitem[1995]{Lau95}
Lauer, T.R., Ajhar, E.A., Byun, Y.-I., et al. 1995, AJ, 110, 2622
\bibitem[2002]{Lau02}
Lauer, T.R., Gebhardt, K., Richstone, D., et al. 2002, AJ, 124, 1975
\bibitem[1974]{Luc74}
Lucy, L.B. 1974, AJ, 79, 745
\bibitem[1998]{Mag98}
Magorrian, J., Tremaine, S., Richstone, D., et al. 1998, AJ, 115, 2285
\bibitem[1999]{Mar99}
Martel, A.R., Baum, S.A., Sparks, W.B., et al. 1999, ApJS, 122, 81
\bibitem[2001]{Mcl01}
McLure, R.J., \& Dunlop, J.S. 2001, MNRAS, 327, 199
\bibitem[2004]{Mcl04}
McLure, R.J., Willott, C.J., Jarvis, M.J., et al. 2004, MNRAS, 351, 347
\bibitem[2001]{Mer01}
Merritt, D., \& Ferrarese, L. 2001, in {\it The central kpc of starbursts and AGN: the La Palma connection"}, proceedings of 
a conference held in Los Cancajos, La Palma, Spain, eds. J.H. Knapen, J.E. Beckman, I. Shlosman, and T.J. Mahony,
ASP Conf. Series, 249, 335  
\bibitem[2004]{Mer04}
Merritt, D. 2004, in Coevolution of Black Holes and Galaxies, ed. L.C. Ho (Cambridge University Press), 264
\bibitem[2002]{Mil02}
Milosavljevi{\'c}, M., Merritt, D., Rest, A., \& van den Bosch, F.C. 2002, MNRAS, 331, L51
\bibitem[1992]{Mor92}
Morganti, R., Ulrich, M.-H., \& Tadhunter, C.N. 1992, MNRAS, 254, 546
\bibitem[1997]{Mor97}
Morganti, R., Parma, P., Capetti, A., Fanti, R., \& de Ruiter, H. R. 1997, A\&A, 326, 919
\bibitem[1987]{Par87} 
Parma, P., Fanti, C., Fanti, R., Morganti, R., \& De Ruiter, H.R. 1987, A\&A, 181, 244
\bibitem[1999]{Par99}
Parma, P., Murgia, M., Morganti, R., et al. 1999, A\&A, 344, 7
\bibitem[2003]{Par03} 
Parma, P., de Ruiter, H.R., Capetti, A., et al. 2003, A\&A, 397, 127
\bibitem[1999]{Pel99}
Pellegrini, S. 1999, A\&A, 351, 48
\bibitem[2002]{Rav02}
Ravindranath, S., Ho, L.C., Filippenko, A.V. 2002, ApJ, 566, 801
\bibitem[2001]{Res01}
Rest, A., van den Bosch, F.C., Jaffe, W., et al. 2001, AJ, 121, 2431
\bibitem[1972]{Ric72}
Richardson, W.H. 1972, J.Opt.Soc.A., 62, 52
\bibitem[1972]{San72}
Sandage, A. 1972, ApJ, 178, 25
\bibitem[1980]{Smi80}
Smith, H.E., \& Spinrad, H. 1980, PASP, 92, 553
\bibitem[2004]{Tru04}
Trujillo, I., Erwin, P., Asensio Ramos, A., \& Graham, A.W. 2004, AJ, 127, 1917
\bibitem[1999]{Van99}
van der Marel, R.P. 1999, AJ, 117, 744
\bibitem[1999]{Ver99}
Verdoes Kleijn, G.A., Baum, S.A., de Zeeuw, P.T., \& O'Dea, C.P. 1999, AJ, 118, 2592
\bibitem[1999]{Wan99}
Wandel, A. 1999, ApJ, 519, 39
\bibitem[1995]{Wil95}
Wilson, A.S., \& Colbert, E.J.M. 1995, ApJ, 438, 62
\bibitem[2004]{Woo04}
Woo, J.-H., Urry, C.M., Lira, P., van der Marel, R.P., \& Maza, J. 2004, ApJ, 617, 903

\end{thebibliography}
\end{document}